\documentclass[12pt]{article}
\usepackage{amsfonts}
\usepackage{amssymb}
\newtheorem{rem}{Remark}
\newtheorem{lem}{Lemma}

\begin{document}

\centerline{\bf P.Grinevich \footnote{P.Grinevich, Landau Institute
for Theoretical Physics, Moscow; e-mail pgg@landau.ac.ru},
S.Novikov\footnote{S.Novikov, University of Maryland, College Park
and Landau Institute for Theoretical Physics, Moscow, e-mail
novikov@ipst.umd.edu} }

\vspace{0.5cm}

\centerline{\Large Singular Finite-Gap Operators and Indefinite
Metric. II\footnote{First version of this work can be found on the
website http://arxiv.org/abs/0903.3976 New results are added
 at the end of the article. New references are included after discussion with several
 colleagues. Especially fruitful discussion with I.Krichever helped us to improve
 our article.}}

\vspace{1cm}

{\bf Abstract}. {\it Many ''real'' inverse spectral data for
periodic finite-gap operators (consisting of Riemann Surface with
marked ''infinite point'', local parameter and divisors of poles)
lead to operators with real but singular coefficients. These
operators cannot be considered as self-adjoint in the ordinary
(positive) Hilbert spaces of functions of $x$. In particular, it is
true for the special case of  Lame' operators with elliptic
potential $n(n+1)\wp(x)$ where eigenfunctions were found in XIX
Century by Hermit. However, such Baker-Akhiezer (BA) functions
present according to the ideas of  works \cite{KN,GN},  right analog
of the Discrete and Continuous Fourier Bases on Riemann Surfaces. It
turns out that these operators for the nonzero genus are symmetric
in the indefinite inner product, described in this work. The analog
of Continuous Fourier Transform is an isometry in this inner
product. Its image in the space of  functions of the variable $x\in
R$ is described.}

{\bf Introduction}

Broad family of the so-called ''Baker-Akhiezer'' (BA) functions on
Riemann surfaces were invented  since 1974 when periodic finite gap
solutions were found for the famous KdV equation.

 They were used for the
solution of periodic problems for KdV, KP and other systems of
Soliton Theory like NLS, SG, for many  Completely Integrable
Hamiltonian Systems. The Spectral Theory of ''finite-gap'' periodic
1D and 2D Schrodinger Operators was developed since 1974 based on
the Analysis on Riemann Surfaces.  It was found in 1987  that some
BA functions generate construction of analogs of the Laurent-Fourier
decomposition for functions and tensor fields on Riemann surfaces
(the Krichever-Novikov Bases and Algebras \cite{KN}). They were used
for the multi-loop operator quantization of Closed Bosonic Strings
(i.e. for genus more than zero.) Another ideas similar to some sort
of Harmonic Analysis with spectral parameter on Riemann Surfaces and
useful here, were developed for other goals in the works \cite{GO}.

The present authors observed in the Appendix to the work \cite{GN}
that these constructions lead also to the analog of  continuous
Fourier  Transform. The present work is  direct continuation of
\cite{GN}. It was motivated by the following

 {\bf Problem}: Consider  one-dimensional  Lame'
 Operator $L=-\partial_x^2+u(x)$ whose potential $u$ is equal to the
 $n(n+1)$-times Weierstrass elliptic function $\wp $ with poles on real
 line. {\bf Does it have any reasonable spectral theory on the whole real line?}
 We need to answer this question because our analog of continuous
 Fourier Transform is based
  exactly on the singular Hermit eigenfunctions of this operator
 in the simplest nontrivial elliptic case.

  Let us remind here that 150 years ago Hermit
 found all family of formal eigenfunction for this operator. In fact it consists of the
 ''Bloch-Floquet'' eigenfunctions in modern terminology. However they are singular
 on the line and do not serve any spectral problem in Hilbert space.
 Hermit used only those of them who belong to the discrete spectrum on the finite interval
 $[0T]$ between the neighboring singularities,  needed for the Lame' problem.
  No spectral interpretation of
singular eigenfunctions for the spectral theory on the whole line was known.

We found indefinite inner product associated with this problem. This
is our main result but the exposition is more general: We
constructed indefinite inner products associated with BA functions
and non-selfadjoint ''algebraic'' periodic operators with
Bloch-Floquet function meromorphic on  Riemann Surfaces of finite
genus like in the finite-gap theory.

{\bf We assume in this work, that all finite-gap operators are
periodic in the real variable $x\in R$.} It is very likely, that our
main results   are valid for generic finite-gap $x$-quasiperiodic
potentials, but this extension may  lead to additional analytical
difficulties.

\begin{rem} {\it Singular Bloch-Floquet eigenfunctions are known also for
the $k+1$-particle Moser-Calogero operator with Weierstrass elliptic
pairwise potential if coupling constant is equal to $n(n+1)$. They
form (if  center of mass is staying) a $k$-dimensional complex
algebraic variety. The Hermit-type result is not obtained here yet
for $k>1$: no one function was constructed until now serving the
discrete spectrum in the bounded domain inside of  the poles. Our
case corresponds to $k=1$. We believe that for all $k>1$ this
algebraic family of eigenfunctions also serves spectral problem in
some indefinite inner product in the proper space of functions
defined in the whole space $R^k$, similar to the case $k=1$.}
\end{rem}

\vspace{2cm}

{\bf Chapter 1. Canonical contours and Inner Product of BA
functions.}

Let a nonsingular complex algebraic curve (Riemann surface) $\Gamma$
be given with selected point $P=\infty\in \Gamma$,  local coordinate
$z=k^{-1}$ near $P$ such that $z(P)=0$. We fix also ''divisor''
$D=\gamma_1+...+\gamma_g$ on $\Gamma$ and construct  standard BA
function $\Psi_D(x,z),z\in \Gamma,$ meromorphic in the variable $z$,
with first order poles in the points $\gamma_j\in \Gamma$ and with
asymptotics $\Psi=\exp\{ikx\}(1+O(k^{-1}))$. We define a
differential 1-form $d\mu$ with asymptotics $d\mu=dk+regular$ near
the infinity $P$, and  zeroes of the form  $(d\mu)=D+D^*$ such that
$$D+D^*\sim K+2P$$ Here the sign $\sim$ means the so-called ''linear
equivalence'' of divisors in Algebraic Geometry, $K$ means the
divisor of differential forms. So the divisor $D^*$ is completely
determined by the divisor $D$. A ''dual'' BA function (1-form)
$\Psi^*_D(x,z)$ was invented long ago by Krichever. It was actively
used in the joint works \cite{KN} and has asymptotics
$\Psi^*_{D}(x,z)=\exp\{-ikx\}(1+O(k^{-1}))$ with divisor $D^*$. A
Dual BA form is $\Psi^*d\mu$. So we have
$$\Psi^*_D(x,z)= \Psi_{D^*}(-x,z)$$ as a scalar BA function. One
should multiply it by the form $d\mu$ to get a dual 1-form.

Both functions  $\Psi_D(x,z)$,  $\Psi^*_D(x,z)$ are  meromorphic in
$x$.

{\bf The Canonical Contours $\kappa_c$} we define by the equation
$p_I=c$ where $dp$ is meromorphic (second kind) differential form
such that $dp=dk+regular$ near $P=\infty$, and $\oint_{\gamma}dp\in
R$ is purely real for all closed paths  $\gamma\subset\Gamma$
avoiding the point $P$. So the imaginary part of $p$ is an
one-valued function $p_I$. {\bf The Canonical Contour is canonically
oriented by the one-valued real function $p_I$ on the oriented
manifold $\Gamma$. The Special Canonical Contour corresponds to the
case $c=0$ such that the expression $\exp\{ikx\}$ is bounded for
real $x\in R$ and $k\rightarrow\infty$ along the contour.}
\begin{rem}
The finite-gap operator, constructed by the curve $\Gamma$ is
$x$-periodic with the period $T$ if and only if the function $e^{iTp}$
is single-valued in  $\Gamma$, or equivalently, if all periods of $dp$
have the form:
$$
\oint_sdp = \frac{2\pi}{T}n_s, \ \ n_s\in\mathbb{Z},
$$
where $s$ in an arbitrary closed contour.
\end{rem}

\begin{rem}{\it In the work \cite{GN} we especially considered the case where
our divisor $D$ is equal to $D=gP=g\infty$ where $g$ is genus of
$\Gamma$. In this case we proved important ''Multiplicative
Property'' of our ''Fourier'' BA basis:
$$\Psi_{g\infty}(x,z)\Psi_{g\infty}(y,z)=L_g\Psi_{g\infty}(x+y,z)$$
where $L$ is a linear differential operator in $x$ with coefficients
independent on $z\in \Gamma, L=\partial_x^g+...$-- see\cite{GN}.
This construction extends the construction \cite{KN} of the discrete
Fourier bases done in the late 1980s for the needs of the Bosonic
(closed) String Theory. The multiplicative properties of Fourier
type series and transform are important in the Nonlinear Problems
like String Theory. This specific case is not much different from
others in the purely linear Harmonic Analysis discussed in the
present work. Poles of $\Psi$ in the variable $x$ necessary appear
in this case, so our inner products are indefinite--see below.
Especially effective formulas for operator $L_g$ were obtained in the work
\cite{B}.}
\end{rem}

 Let us  define {\bf a $C$-linear Inner Product} of
smooth functions on the canonical contour $\kappa_c\subset \Gamma$
depending on the choice of divisor $D$ and generated by the basis of
functions $\Psi_D(x,z)$ restricted to the canonical contour
$\kappa_c$.

{\bf Statement.} For the basic BA functions we have  {\bf The
Orthogonality Relations on Riemann Surface, i.e. on the contour
$\kappa_c\subset\Gamma$}:

$$(\Psi_D(x,z),\Psi_D(y,z))_{\kappa_c}=\int_{\kappa_c}\Psi_D(x,z)\Psi_{D^*}(-y,z)d\mu(z)=$$
$$=2\pi\delta(x-y)$$
The exact meaning of this statement and of the $\delta$-function
depends on the functional classes where this product is
well-defined. For the specific real nonsingular ''self-adjoint''
case with positive inner product such relation was first found in
the work \cite{Kr}--see formula 12. {\bf We are going to use this
type of relations for the singular operators where inner product is
not positive. The most important case for us is the case of Special
Canonical Contours $c=0$}. Let us present sketch of the proof:

 The form at the right-hand side is
holomorphic in the variable $z$. Therefore this integral does not
depend on $c$. If $x>y$, this integral vanishes as
$c\rightarrow+\infty$. Equivalently, for $x<y$ this integral
vanishes as $c\rightarrow-\infty$, therefore
\begin{equation}
\label{eq:z-orth}
(\Psi_D(x,z),\Psi_D(y,z))_{\kappa_c}=0 \ \ \mbox{for} \ \ x\ne y.
\end{equation}
If we modify the integrand outside a neighbourhood of the point $P$,
the resulting integral is the same up to a regular function of $x$,
$y$. Let us expand the functions $\Psi_D(x,z)$,  $\Psi^*_D(x,z)$ near
the point $P$:
$$
\Psi_D(x,z)=e^{ikx}\left[1+\frac{\phi(x)}{k}+O\left(\frac{1}{k^2} \right) \right]
$$
$$
\Psi^*_D(x,z)=e^{-ikx}\left[1+\frac{\phi^*(x)}{k}+O\left(\frac{1}{k^2}
  \right) \right]
$$
$$
(\Psi_D(x,z),\Psi_D(y,z))_{\kappa_c}=\int_{-\infty}^{+\infty}
e^{ik(x-y)} \left[1+\frac{\phi(x)+\phi^*(y)}{k}\right]dx + \mbox{regular function}=
$$
$$
=2\pi\delta(x-y) + \pi i \, \mbox{sgn}(x-y) [\phi(x)+\phi^*(y) ]+ \mbox{regular function}.
$$
For $x=y$ the integrand has no essential singularities and only one
first-order pole at $P$ with the residue $\phi(x)+\phi^*(x)$.
Therefore $\phi^*(x)=-\phi(x)$, and
$$
(\Psi_D(x,z),\Psi_D(y,z))_{\kappa_c}=2\pi\delta(x-y) + \mbox{regular function}.
$$
Comparing it with (\ref{eq:z-orth}) we complete the proof.

Now  we consider class of functions $\phi(z)$ on the Special
Canonical Contour $\kappa_0$, such that their {\bf ''BA Fourier
Transform''} is well defined. We interpret them simply as {\bf ''BA
Fourier Components''} of function $\tilde{\phi}(x)$ in our {\bf BA
Fourier basis} $\Psi_{D^*}(-x,z)$ using the integral:
$$\tilde{\phi}(x)=(\sqrt{2\pi})^{-1}\int_{\kappa_c}\phi(z)\Psi_{D^*}(-x,z)d\mu$$

{\bf Statement.} For the selected BA function with bounded
restriction of $\exp\{ikx\}$ to the Special Canonical  Contour
$\kappa_0$ near $P=\infty$,  this integral is  well-defined near
$\infty$ if $\phi(k)=o(k^{-1+\epsilon}),\epsilon>0$.

Proof. The {\bf Inverse BA Fourier Transform} is given by the
formula
$$\phi(z)=(\sqrt{2\pi})^{-1}\int \tilde{\phi}(x)\Psi_D(x,z)dx$$
It leads to the same inner product of transformed functions in the
$x$-space treated simply as ''Collections of BA Fourier Components''
in the previous basis of BA functions $\Psi_D(x,z)$, where $x$ is
considered as an ''index'' numerating the basic vectors:
$$\int\tilde{\phi}_1(x)\tilde{\phi}_2(x)dx=(\phi_1,\phi_2)_{\kappa_c}$$
{\it The space of functions in the $x$-space obtained by this  BA
Fourier Transform will be especially discussed below for the
important real case, after the proper definition of the space of
functions $H_{D,\kappa_0}$ on corresponding contour in $\Gamma$}. We
describe its construction for the most important hyperelliptic case
in the Appendix 2.

 For the same BA functions treated as basis in $x$-space,
  we obtain a formula
\begin{equation}
\label{eq:orth2}
(\Psi_D(x,z),\Psi_D(x,w))_{x}=[\int_x\Psi_D(x,z)\Psi_{D^*}(-x,w)dx]d\mu
 = 2\pi\delta(z,w)
\end{equation}
 $=0,z\neq w$. Here both points $z,w\in \Gamma$ belong to our selected contour
 $\kappa_c$. We assume that these points are nonsingular on this
 contour. We assume that $\delta$ is an one-form in the variable $w$.

  The case of critical contour corresponding to the critical
 values of the real function $p_R$ should be considered separately.
 {\bf This formula is meaningful locally only if our BA functions do not contain
 poles for $x\in R$. It is meaningful globally if our picture is
 periodic in $x\in R$}, so we have no concentration of poles near
 $x\rightarrow \pm\infty$. We postpone to the next work extension of
 our results to the quasi-periodic finite-gap case.

Let us discuss, for which classes of functions our BA Fourier
Transform is well-defined. It depends on the divisor $D$ and on the
geometry of contour $\kappa_c$: {\bf Does our BA function contain
poles? Does divisor contain infinite point or not? Is our contour
critical?}

We postpone the last question.

For the case of BA function with poles we invent following  rule:
{\bf All integrals above taken along the line $x\in R$, should be
taken avoiding pole $x_0$ in the upper half-plane $x+i\epsilon,
\epsilon >0$.} In order to prove that our inner products written as
integral along the $x$-axis, are well-defined, we prove following:

\begin{lem} The expression $\Psi_D(x,z)\Psi_{D^*}(-x,w)$ has
residue equal to zero in every pole $x_0\in R$ as a meromorphic
function of the complex variable $x$ in the small strip around the
real line.

\end{lem}

The proof follows immediately from Lemma~\ref{lemma:mer_omega} and
formula~(\ref{eq:omega_der}) below. We added Appendix 1 to make this
proof fully rigorous. In the case of first order poles in the point
$x=0$ both our BA functions have a form like $\alpha/x+O(x)$.
Product of such expressions obviously has zero residue.

 So  the integral defining inner product does not depend on
the contour around the pole.

{\bf Real Algebraic Curves.} Let our data consisting of algebraic
curve $\Gamma$ with selected point $P=\infty$ and local parameter
$z=k^{-1}$ near $P$, be real now. It means precisely that an
anti-holomorphic involution is defined
$$\tau:\Gamma\rightarrow \Gamma, \tau^2=1$$ such that $\tau(P)=P$ and
$\tau^*(k)=\bar{k}$.

 Our differential $dp$ is such that
$\tau^*(dp)=d\bar{p}$. We define $p_I$ such that $\tau^*(p_I)=-p_I$,
so the level $\kappa_0=(p_I=0)$ is invariant under $\tau$:
$$\tau:\kappa_0\rightarrow\kappa_0$$
and differentials $dk, dp$ are real on $\kappa_0$.

 Let us point out
that our contour $\kappa_0$ contains all set of fix-points
$$Fix_{\tau}\subset \kappa_0$$ where $z\in Fix_{\tau}$ means
$\tau(z)=z$.  Following simple geometric statement is useful to
clarify relationship between our constructions and some results of
the late 1980s (see\cite{DNa}) about nonsingular real solution to
the KPI system with Lax operator $i\partial_x+\partial_y^2+u(x,y)$:

\begin{lem}
For anti-holomorphic involution $\tau$ the fix-point set
$Fix_{\tau}$ coincides with canonical contour $\kappa_0$ if and only
if $Fix_{\tau}$ divides $\Gamma$ into two parts
$\Gamma=\Gamma_+\bigcup\Gamma_-$.
\end{lem}
Proof of this lemma easily follows from the obvious fact that
$\kappa_0$ certainly divides $\Gamma$ but its smaller  part never
does. We assume that $P\in Fix_{\tau}$.

 We choose
divisor $D$ such that $\tau(D)=D^*$ or
$$D+\tau(D)\sim K+2P$$ where $K$ is divisor of differential
forms. So we have $\tau^*(d\mu)=d\bar{\mu}$.

In this case we define a {\bf Hermitian (or sesqui-linear) possibly
indefinite Inner Product} for the basic BA functions on the contour
$\kappa_0$ by the formula
$$<\Psi_D(x,z),\Psi_D((y,z)>_{\kappa_0}=(\Psi_D(x,z),\bar{\Psi}_D(y,\tau(z)))_{\kappa_0}=$$
$$=\int_{\kappa_0}\Psi_D(x,z)\bar{\Psi}_{\tau D}(y,\tau z)d\mu(z)$$
where the integral above is taken with respect to the canonical
orientation of the contour $\kappa_0$.

 We take into account here
that $\bar{\Psi}_{\tau D}(y, \tau z)$ is meromorphic in the variable
$z$, has poles in $\tau D$ and asymptotics
$\exp\{-iky\}(1+O(k^{-1}+...)$ near $P$ for $y,k\in R$. So for the
''real'' variables it coincides with our $C$-linear expression
above.

 In the $x$-space we have following
inner product of basic BA functions:
$$<\Psi_D(x,z),\Psi_D(x,w)>_x=\int_x [\Psi_D(x,z)\bar{\Psi}_{\tau
D}(\bar{x},w )d\mu] dx $$ Let us point out that $$\bar{\Psi}_{\tau
D}(\bar{x},w)= \Psi_{\tau D}(-x, w)$$ for the real values of the
variables $k,x$. It is meromorphic in $x$. So the residue in the
$x$-pole is equal to zero for the product
$\Psi_D(x,z)\bar{\Psi}_{\tau D}(\bar{x}, \tau(z))$ in the integral
because it is the same as in the Lemma 1 above.

 We are coming to the following

\begin{lem}  1.The hermitian inner product above  on the contour $\kappa_0$
is positively defined if and only if $\kappa_0=Fix_{\tau}$, and the
form $d\mu$  is positive on the contour $\kappa_0$. 2. The hermitian
inner product in the $x$-space is well-defined avoiding every pole
of $\Psi$ in the upper half-plane in $x$. It is positive if and only
if our BA function $\Psi_D(x,z)$ does not have poles on the real
line $x$.
\end{lem}

The statement 1 makes sense because our form $d\mu$ is real on this
contour. We have $\tau(z)=z$ for $z\in Fix_{\tau}$, and upper part
$\Gamma^+$ of $\Gamma$ induces natural orientation of the contour
$\kappa_0$. It is interesting to compare this result with
\cite{DNa}. The statement 2 is crucial for our work, so we present a
full proof in the Appendix 1 using what we call The
Cauchy-Baker-Akhiezer Kernel. This quantity is borrowed from the
work \cite{GO}, but some additional improvements are needed here.
Besides that, no full proof was presented in the work \cite{GO}.

{\bf There are following sources for the violation of positivity of
the inner product} on the contour $\kappa_0$:

1.  $\kappa_0\neq Fix_{\tau}$. We have here $\tau(z)\neq z$ for $z$
outside of fix-point set. Such inner product is always indefinite.

2. $Fix_{\tau}=\kappa_0$ but the divisor $D$ is chosen such that
$d\mu$ has different signs  on some components (see chapter 2).

{\bf Only poles of $\Psi$ on the real line $x$ are responsible for
the non-positivity of the inner product in the $x$-space. This is
central part of our work.}

We are going to consider this picture in more details for the
important hyperelliptic case in the next chapter.

\vspace{2cm}

{\bf Chapter 2.  The indefinite Inner Product for Hyperelliptic
Riemann Surfaces.  Schrodinger Operators with singular potential.}

Consider now the most important case of nonsingular Hyperelliptic
Riemann Surfaces $\Gamma$ associated with second order periodic
operators $L$: Let $\Gamma$ is presented in the form
$$w^2=(u-u_0)\times ...(u-u_{2g})=R(u)$$ where typical point (except
branching points) is written as $\gamma=(z,\pm)$. We take branching
point $P$ with $u=\infty$ as our ''infinity'' with local coordinate
$k^{-1}=u^{-1/2}=z$. Every generic divisor $D=\gamma_1+...+\gamma_g$
defines a Baker-Akhiezer function $\Psi_D(x,z)$ with standard
analytic properties described above. It satisfies to the equation
$$L\Psi=(-\partial_x^2+U(x))\Psi(x,z)=u(z)\Psi(x,z)$$
Our requirement is that the potential $U(x)$ is periodic
$U(x+T)=U(x)$ for real $x$. From the finite-gap theory we know that
necessary and sufficient condition to have real nonsingular
potential $U(x)$ (we call it a {\bf Canonical Inverse Spectral
Conditions}) consists of two parts:

1.The Strong  Reality Condition for $\Gamma$: all branching points
$u_j$ are real and distinct. Let $u_0<u_1<...<u_{2g}$.

2.The divisor $D$ is {\bf Proper} i.e. such that
$\gamma_k=(\alpha_k,+)$ or $\gamma_k=(\alpha_k,-)$ where
$u_{2k-1}\leq \alpha_k\leq u_{2k}, k=1,2,...,g$ (exactly one divisor
point is located in every $a$-cycle).

There are two commuting anti-holomorphic  involutions $\tau_{\pm}$
of the Riemann Surface $\Gamma$ where
$\tau_{\pm}(u,+)=(\bar{u},\pm)$. Let $\tau_+=\sigma,\tau_-=\tau$.
Our contour $\kappa_0$ is equal to $Fix_{\tau}$. It coincides with
union of spectral zones. The set $Fix_{\sigma}$ coincides with union
of spectral gaps:

The union of our $a$-cycles $a_k$ form the finite part of fix-point
set for the anti-involution $\sigma (p)=p$. Their projection on the
$u$-line $u\in R$ coincide with  finite ''Gaps''
$[u_{2k-1},u_{2k}],k=1,2,...,g$, in the Spectral Theory of  operator
$L$ in the Hilbert Space $L_2(R)$ of the square-integrable real or
complex-valued functions on the real line. So we have $\sigma (D)=D$
and $\tau D=D^*$ where $D+D^*=K+2P$.

We know however that inverse spectral data lead to the real
operators $L$ in other {\bf Non-Canonical Real Cases}:

1.The Riemann Surface $\Gamma$ is Real.  It simply means that the
set of branching points  $u_j,j=0.1,...,u_{2k}$, is invariant under
the anti-involution $u\rightarrow \bar{u}$.

2.The Divisor of Poles $D$ should be such that $\sigma (D)=D$ but
not necessarily like in the Canonical Case.

If these conditions are satisfied, then the potential $U(x)$ is
real. However, this potential is singular . Otherwise, it would be
self-adjoint in the positive Hilbert Space which is impossible. So
it is singular in all non-canonical real cases. We call our data
{\bf Real Semi-Canonical} if Riemann Surface satisfies to the Strong
Reality Condition but the divisor $D$ is nor Proper. In particular,
our contour $\kappa_0$ coincides with fix-point set $Fix_{\tau}$.
The potential $U(x)$ has poles in all real non-canonical cases, and
spectrum is real in the semi-canonical case.

Orientation of $\kappa_0$ in the Real  case is defined by the domain
$p_I\geq 0$ and orientation of $\Gamma$. For such spectral curves
\begin{equation}
\label{eq:dp} dp= (u-p_1)\times ...(u-p_{g})du/\sqrt{(u-u_0)\times
...(u-u_{2g})},
\end{equation}
where all $p_k$ are real and $p_k\in [u_{2k-1},u_{2k}]$.

Let not all branching points are real: there are complex adjoint
pairs between them. In this case we have $Fix_{\tau}$ essentially
smaller than the contour $\kappa_0$. So our operator $L$ is
singular. It has complex spectrum equal to the projection of the
contour $\kappa_0$ on the complex $u$-line.

Such operators are symmetric in the Indefinite Inner Product given
by the formulas presented in the Chapter 1.

Using previous results, we are coming to the following

{\bf  Theorem}: {\it 1.Let Riemann Surface and divisor $D$ are real
and finite correspondingly.
 The form
 $$d\mu=(u-\gamma_1)\times...(u-\gamma_g)du/\sqrt{R(u)}, u=z^{-1/2},$$
  is real, nonzero and has a
 well-defined sign in every component. 2.The set $\kappa_0$  is   the spectrum of operator $L$
 in some space of functions $\tilde{H}_L$ of the variable $x$ depending on the poles of
 operator $L$
  (see description below
 in the generic case).
  The Inverse BA Fourier Transform defined above
  isomorphically maps the space $\tilde{H}_L$ on the space $H_{D,\kappa_0}$
  of functions on the contour
   $\kappa_0$ and back. An Indefinite  Inner Product $$< a,b >_{\kappa_0}=
   \int_{\kappa_0}a(z)\bar{b}(z)d\mu(z
   )$$ is defined
    in the space $H_{D,\kappa_0}$. It is isomorphic   to the direct sum of
     ordinary spaces  of
functions in the components of $\kappa_0$, taken with sign provided
by the form $d\mu$ and orientation of the contour $\kappa_0$; The
decay in the infinite component is also specified as above.
 The linear operator $L=-\partial_x^2+U(x)$ is symmetric in the space $\tilde{H}_L$,
and corresponding ''Direct and Inverse BA Fourier Transform''
defined in the previous paragraph, is isometric corresponding to
these indefinite inner products. 3.In the case where some  divisor
points are infinite $\gamma_g=\infty$,
 the form $d\mu$ is holomorphic.}

\begin{rem}
From (\ref{eq:dp}) it follows that the sign of $d\mu$ on real ovals
of $\tau$ with respect to the orientation of $\kappa_0$ coincides
with the sign of $dp/d\mu$, or, equivalently, with the sign of the ratio:
$$
(u-p_1)\ldots(u-p_g)/(u-\alpha_1)\ldots(u-\alpha_g).
$$
\end{rem}

\begin{rem}
As one might see below, we do not describe the exact completion of
the spaces $\tilde{H}_L$ and $H_{D,\kappa}$. So our result is
incomplete in terms of Modern Functional Analysis.
\end{rem}

\begin{rem}
If $r$ point of divisor $D$ are equal to $\infty$
(i.e. $D=r\infty +(\alpha_1,\pm)+...+(\alpha_{g-r},\pm)$), we have
$$d\mu=(u-\alpha_1)\times
...(u-\alpha_{g-r})du/\sqrt{R(u)}$$ The special case $r=g$ all
divisor is concentrated in the point $\infty$. This case was
especially considered as a right analog of the ordinary Fourier
Transform: It has Remarkable Multiplicative Properties.
\end{rem}

Proof. Our Theorem immediately follows from the results of Chapter 1
and Appendix 1.

 {\bf Description of the Space $\tilde{H}_L$ for generic
 singular finite-gap real $x$-periodic operators $L$}:

  Let us consider real singular ''finite-gap'' periodic
 potentials  $U(x)$ with finite number of poles with cyclic order $x_0=0<x_1
<...<x_k<T$ at the circle $[0T]$. In the generic finite-gap case we
have
$$U=2/(x-x_j)^2+O(x-x_j)$$ near every pole $x_j$.  We define following class
of functions $f(x)\in \tilde{H}_{x_0,...,x_k}$ depending on the
position of poles only, by the requirement:
$$f(x)=\alpha_j/(x-x_j)+O(x-x_j)$$ where $f(x)-\alpha_j/(x-x_j)=O(x-x_j)$ is
 $C^{\infty}$-smooth
near the point $x=x_j$. By definition of this class, the operator
$L=-\partial^2+U(x)$ is locally well-defined in it:
$$Lf(x)\in \tilde{H}_{x_0,...,x_k}$$
Describing these spaces, we decompose our space of functions on the
line $R$ into the ''direct integral'' along the Bloch-Floquet
multiplier $\varkappa$. All BA (i.e. Bloch-Floquet) eigenfunctions
for all points $z\in \kappa_0$ locally belong to this classes for
some $\varkappa$ as it follows from the formulas for BA functions
obtained in the theory of finite-gap operators.

{\bf Question:}Are there any $L$-dependent linear relations between
these residues for singular finite-gap periodic operators?

 It
means that some linear subspace $C^q(L, \varkappa)\subset C^k$
should be chosen in the space of residues. It depends on the
Bloch-Floquet multiplier $\varkappa$ and operator $L$. Its dimension
$q=q(L)$ is equal to the number of negative squares in our inner
product for the fixed value of the Bloch Floquet multiplier
$\varkappa$--see the case of higher Lame potentials in the Appendix
2. {\bf Therefore our indefinite space for every $\varkappa$
consists of all functions $f\in F_{x_0,...,x_k}(\varkappa)$ with
first order zero in all poles $x_j$ plus finite-dimensional space
$C^q(L,\varkappa)\subset C^k$}:
$$\tilde{H}_{L,\varkappa}=C^q(L,\varkappa)\bigoplus F_{x_0,...,x_k}(\varkappa)$$
 The total space
$\tilde {H}_L\subset \tilde{H}_{x_0,...,x_k}$ we realize as a direct
integral of the spaces $\tilde{H}_{L,\varkappa}$ over the circle.

{\bf We proved in many cases that in fact $k=q$, i.e. there are no
relations between these residues. In particular, it means that the
space $\tilde{H}_L$ depends on the poles only (see Appendix 2).
According to our Conjecture this statement is true for all
finite-gap periodic potentials with generic poles. Natural extension
of this result for periodic potentials with isolated poles of the
form $n(n+1)/2+O(x)$
 is also described below}.

 Take any smooth functions $a(z)$ on the contour
$\kappa_0$ properly decreasing in the infinite component for
$z\rightarrow 0$ or $k\rightarrow \infty$. We can see that its BA
Fourier Transform
$$f(x)=\tilde{a}(x)=1/\sqrt{2\pi}\oint_{\kappa_0}a(z)\Psi_{D^*}(x,z)
d\mu$$  belongs to the  class described above with some proper decay
at $|x|\rightarrow \infty$. We  define global behavior and
completion of the space in the Appendix 2 using operators on the
circle $[0T]$ with periodic (quasi-periodic) boundary conditions at
the ends  with unimodular Bloch-Floquet multipliers
$|\varkappa|=|\exp\{ip(z)T\}|=1$ on the contour $\kappa_0$ where
$p=p_R$.

 {\bf
Examples: The case $g=1$.} There are two different cases here:

{1.The Hermit-Lame' Operator.} Consider real elliptic curve $\Gamma$
with genus $g=1$ and real branching points $u_0,u_1,u_2,\infty$.
{\bf So this case is real semicanonical}. We assume that our divisor
$D=\gamma$ coincides with $P=\infty$. The Baker-Akhiezer Function
$\Psi_D(x,z)$ here was found by Hermit simply as some basis of
solutions for all values of spectral parameter $u(z)=z^{-2}$.
Operator $L$ here is the Lame Operator; it has periodic potential
$U(x)=2\wp(x)$ with poles like $2/x^2$ on the real line $x$ in the
points $nT,n\in Z$. Classical XIX Century people considered this
operator on the interval $[0T]$. They obtained this operator (by the
separation of variables in Jacoby coordinates) from the
Laplace-Beltrami operator on the 3-axis ellipsoida. They needed its
spectrum on the interval $[0T]$ with zero boundary conditions at the
ends $0,T$. {\bf It is normally called in the literature ''The
Dirichlet Spectrum'' but we call it The Hermit Spectrum for
finite-gap operators}. Let us make some useful methodological
remarks about the comparison of Hermit (Dirichlet) Spectrum and
periodic  spectrum of Nonsingular Shifted Periodic Operator
$L_{i\omega}$ on the circle $0T$ and with spectrum on the whole line
$R$--{\bf The One-Gap Operator.}

{\bf Statement.} The Hermit spectrum of Lame operator $L$ is simple.
It consists exactly of eigenvalues $\lambda_s, s\in Z^+$, which are
the double-degenerate eigenvalues of the  shifted nonsingular
periodic operator $L_{i\omega}$ on the circle. Here $2i\omega$ is an
imaginary period of the function $\wp(x)$.

{\it As we know, the potential $2\wp(x+i\omega)$ plays fundamental
role in the Theory of Solitons: this function defines the Traveling
Waves (''Soliton Lattices'') for the KdV Equation as it was found in
XIX Century by Korteweg and De Vries:} Consider the corresponding
Lax Operator
$$L_{i\omega}=-\partial_x^2+2\wp(x+i\omega)$$
The theory of periodic problem for KdV started in 1974 from the
''soliton derivation and understanding'' of the spectrum of this
operator. In particular, its non-degenerate eigenvalues satisfying
to the $\pm$-periodic boundary conditions
$$L_{i\omega}\psi(x)=\lambda\psi(x),\psi(x+T)=\pm\psi(x)$$
are exactly the finite branching points $u_0,u_1,u_2$ of  Riemann
surface $\Gamma$. It has also infinite number of double-degenerate
eigenvalues $\lambda_s$ for $u> u_2$, for the same boundary
conditions. It is exactly the set of all Hermit eigenvalues for the
nonshifted operator $L$. They coincide with extremal points of the
half-trace $S(u)$ of the monodromy matrix $\hat{T}(u)$:
$$1/2Tr\hat{T}=S(u), \hat{T}(\psi(x,u))=\psi(x+T,u)$$
 along the minimal real period $T$. We have $-1\leq S(u)\leq 1,u\in
 [u_2,\infty]$,
and $S(u)=\pm 1,S'(u)=0$ for $u=\lambda_s$. In particular,
$\lambda_s>u_2$.

 Proof. The Hermit problem is real, so we look for the real
 eigenfunctions $\psi(x, \lambda_s)=\psi_s(x)$ equal to zero at the
 ends $0,T$. The space of nonsingular solutions to the equation
 $L\psi=\lambda_s\psi$ is one-dimensional, so this function $\psi$
 should be also an eigenvector of the monodromy matrix $\hat{T}$ with real
 eigenvalue. So it is either the double-degenerate point indicated in our statement or
 it belongs to the gap of the spectrum. In the last
 case we point out that BA functions in the gaps always have poles
 at the points $nT$ for the Lame operator. Our statement is proved.

For the points $u=\lambda_s$ monodromy matrix $\hat{T}$ is equal to
$\pm 1$. One of corresponding two eigenfunctions is nonsingular and
equal to zero in the points $nT$ defining the eigenfunction of the
Hermit Spectrum. So it is completely determined by the Riemann
Surface $\Gamma$.

We return now to  spectral theory of the singular Lame operator $L$
on the whole line $x\in R$ and canonical real contour $z\in
\kappa_0$.

 The Space of functions $H_{D,\kappa_0}$ is a direct sum
of 2 spaces
$$H=H_0\bigoplus H_{\infty}$$ Here $H_0$ consists of functions on
the compact circle $c_1\subset \Gamma$ located over the spectral
zone $[u_0,u_1]  $  (the finite zone of spectrum). The second
subspace $H_{\infty}$ consists of functions on $R\subset \Gamma$
located over the infinite zone of spectrum $[u_2,\infty]$ and
homeomorphic to $R=S^1 \backslash\infty$. They have  specific
asymptotic at infinity indicated above in the chapter 1.

{\bf Statement: The Indefinite  Inner Product in the space
$H_{\infty, \kappa_0}=H_0\bigoplus H_{\infty}$ is positive at
$H_{\infty}$ and negative at $H_0$.}

 Proof. We have in this case
$d\mu=dz/\sqrt{u-u_0)(u-u_1)(u-u_2)}$, and orientation of the
contour $\kappa_0=c_1\bigcup c_{\infty}$ is such that $d\mu_{c_1}<0,
d\mu|_{c_{\infty}}>0$. This statement is proved.

 For comparison good
to consider the ''selfadjoint'' case such that
$\gamma'=(\alpha',\pm)$ where $\alpha'\in [u_1,u_2]$ is located in
the finite gap. In this case we have $$d\mu'=(u-\alpha')
du/\sqrt{(u-u_0)(u-u_1)(u-u_2)}$$ So we have
$d\mu'=(u-\alpha')d\mu$. Taking into account that the function $p_I$
is the same in both cases, we see that the factor $(u-\alpha')$ has
opposite signs in the gaps $c_1$ and $c_{\infty}$. Therefore in this
case the Inner Product is positive (as we knew before).

{\bf 2. The complex branching points.} Let $g=1$ but
$u_0=\bar{u}_1\in C, u_3\in R$ for the Riemann Surface $\Gamma$.
{\bf This case is real but not semicanonical}. We take divisor point
at infinity $D=\infty$. Corresponding potential is also real
function $2\wp(x)$ with pole $2/x^2$ and real period $T$ but
corresponding lattice is romb-like (the complex period is not
orthogonal to the real one, but their lengths are equal to each
other.) Our contour $\kappa_0$ is connected and critical. It
consists of two circles crossing each other transversally in two
real points $u^*,\pm$ on the infinite component over $[u_2,\infty]$.
So the spectrum of operator $L=-\partial_x^2+2\wp(x)$ on the whole
line discussed in this work, contains non-real part where
anti-involution $\tau$ is not identity $Fix_{\tau}\neq \kappa_0$.
The Hermit Spectrum is also well-defined (zero boundary conditions
at the poles $0,T$). It is also equal to the set of real points
$S(u)=\pm 1$, where $S=1/2Tr\hat{T}$ is the trace of monodromy
matrix (which is real for real $x$). All these points are located at
the infinite real zone $[u_3,\infty]$.

{\bf The Standard Fourier Transform} we have for the case of genus
zero: The Riemann Surface $\Gamma$ has 2 branching points
$u_0=0,u_1=\infty$. The spectral zone in $\Gamma$ has one component
$c_{\infty}$ only located over $[0,\infty]$. It is isomorphic to
$R=S^1 \backslash \infty$. The measure $d\mu$ coincides with
standard measure. So our Hilbert Space is exactly
$H=H_{\infty}=L_2(R)$, and inner product is positive. For genus more
than zero we can have positive inner product only for smooth
potentials where divisor points are located in the finite gaps (one
gap--one point).

Return now to the generic singular finite-gap potentials. {\bf In
all cases where our divisor contains infinite point or any point
located at the infinite gap, we have indefinite inner product}. We
can always move this point by some time shift to infinity.

 In all cases where our divisor contains two (or more) points located in the
 same finite gap, we have indefinite inner product.

We can easily describe the  sign corresponding to the cycle
$c_j\subset \Gamma$ located over the zone $[u_{2j-1},u_{2j}]$, i.e.
how it enters  the Indefinite  Space:

Take divisor points $\gamma_s=(\alpha_s,\pm)$, where $\alpha_s\in
[u_{2q_s-1},u_{2q_s}], s=1,...,r$, and $\gamma_{r+k}=\infty$ for all
$k>r$. As we know, it simply coincides with  sign of the expression
$dp/d\mu$ where:
$$d\mu=(u-\alpha_1)\times...(u-\alpha_r)dz/
\sqrt{(u-u_0)\times...(u-u_{2g})}$$ on the cycle $c_k$, taking into
account the orientation of the contour $\kappa_0$ provided by the
function $p_I$ as it was explained in the Chapter 1. {\bf For
example, for $r=0$ (the  case of BA Fourier Transfom with important
Multiplicative Properties based on the Hermit-Lame' potentials
$n(n+1)\wp(x)=U(x)$), the signs corresponding to $c_k$, are
alternating.} Here we have $$d\mu=du/\sqrt{R(u)}$$ We shall discuss
this case in the Appendix 2.

\vspace{2cm}

 {\bf Appendix 1. The Cauchy-Baker-Akhiezer Kernel}

\vspace{1cm}

 Following \cite{GO}, let us define the
Cauchy-Baker-Akhiezer Kernel $\omega(x,z,w)$, $x\in\mathbb{C}$,
$z\in\Gamma\backslash P$, $w\in\Gamma\backslash P$ by the following
analytic properties:
\begin{enumerate}
\item For a fixed $x$ the kernel $\omega(x,z,w)$ is a meromorphic function in $z$
  and a meromorphic 1-form in $w$.
\item For fixed $x,w$ the kernel  $\omega(x,z,w)$ has exactly $g+1$ simple poles
  in $z$ at the points $\gamma_1$, \ldots, $\gamma_g$, $w$.
\item For fixed $x,z$ the kernel $\omega(x,z,w)$ has simple zeroes in $w$
  at the points $\gamma_1$, \ldots, $\gamma_g$, and a simple pole with
  residue 1 at the point $z$. In local coordinates we have
\begin{equation}
\label{eq:omega_diag2}
\omega(x,z,w) = \frac{dw}{w-z} + \mbox{regular terms} \ \ \mbox{as} \
\ w\rightarrow z.
\end{equation}
\item For fixed $x,w$ the kernel $\omega(x,z,w)$ has an essential singularity in
 the variable  $z$ at the point $P=\infty$:
\begin{equation}
\label{eq:omega_as_z} \omega(x,z,w) =
e^{ik(z)x}\left(O\left(\frac{1}{k(z)} \right) \right).
\end{equation}
\item For fixed $x,z$ the kernel $\omega(x,z,w)$ has an essential singularity in
 the variable $w$ at the point $P=\infty$:
\begin{equation}
\label{eq:omega_as_w}
\omega(x,z,w) = e^{-ik(w)x}\left(O\left(\frac{1}{k(w)} \right) \right)dk(w).
\end{equation}
\end{enumerate}

For generic spectral data the kernel $\omega(x,z,w)$ exists and is
unique, by the Riemann-Roch Theorem. The proof is analogous to the
proof of existence and uniquiness for the Baker-Akhiezer function.
Following the idea of the work \cite{GO}, we prove one of the most
important  properties of this Kernel:

 {\bf Fundamental
Lemma}. Following Formula is Valid:
\begin{equation}
\label{eq:omega_der}
\partial_x\omega(x,z,w) = -i \Psi(x,z)\Psi^*(x,w) d\mu(w).
\end{equation}
{\bf Proof.} For a fixed $w$ the right-hand side of
(\ref{eq:omega_der}) has the following analytic properties:
\begin{enumerate}
\item $\partial_x\omega(x,z,w)$ is meromorphic in the variable $z$ on
$\Gamma\backslash P$ and has exactly $g$ simple poles  at the points
$\gamma_1$,\ldots,$\gamma_g$.
\item  $\partial_x\omega(x,z,w) = O(1)e^{ik(z)x}$ as $z\rightarrow P$.
\end{enumerate}
Therefore for a fixed $w$ the expression $\partial_x\omega(x,z,w)$
is proportional to $\Psi(x,z)$. Similarly for a fixed $z$ the
expression $\partial_x\omega(x,z,w)$ is proportional to $\Psi^*(x,w)
d\mu(w)$. Therefore $\partial_x\omega(x,z,w) = c
\Psi(x,z)\Psi^*(x,w) d\mu(w)$. Assuming $z$ and  $w$ are both close
to $P$, we obtain $c=-i$. Our Fundamental Lemma is proved.

\begin{rem}
Let $x=0$. Then the kernel $\omega(0,z,w)$ coincides with the
standard meromorphic analog of Cauchy kernel on the closed Riemann
surfaces (see \cite{Kopp}, \cite{Rodin}).
\end{rem}
\begin{rem}
It is natural to define  Cauchy-Baker-Akhiezer kernel $\omega(\vec
t,z,w)$ depending on all KP times $\vec t=(t_1,t_2,t_3\ldots)$,
$x=t_1$, $y=t_2$, $t=t_3$ (see \cite{GO}). Essential singularities
for the kernel $\omega(\vec t,z,w)$ have the following form:
\begin{equation}
\label{eq:omega_as_z_2}
\omega(\vec t,z,w) = e^{i\sum\limits_{j=1}^\infty t_jk^j(z)}
\left(O\left(\frac{1}{k(z)} \right) \right), \ \ z\rightarrow P,
\end{equation}
\begin{equation}
\label{eq:omega_as_w_2} \omega(\vec t,z,w) =
e^{-i\sum\limits_{j=1}^\infty t_jk^j(w)} \left(O\left(\frac{1}{k(w)}
\right) \right)dk(w), \ \ w\rightarrow P.
\end{equation}
Here we assume that only finite number of variables $t_n$ are
different from 0.
\end{rem}

To stress the dependence of $\omega(\vec t,z,w)$ on the divisor
$D=\gamma_1+\ldots+\gamma_g$, we shall write  $\omega_D(\vec t,z,w)$
if necessary.
\begin{lem}
Denote by $D(\vec t)$ the divisor of zeros for the function
$\Psi_D(\vec t,z)$. Then for any $\vec t'$ we have following
transformation law:
\begin{equation}
\label{eq:omega_tran}
\omega_D(\vec t,z,w) = \frac{\Psi_D(\vec t',z)}{\Psi_D(\vec t',w)}
\omega_{D(\vec t')}(\vec t-\vec t',z,w)
\end{equation}
\end{lem}

The next special case plays leading role in our investigation
because it is the most natural source for the singular operators:

\begin{lem}
Assume, that exactly one point of the divisor $D$ lies at the infinite
points $P=\infty$: $D=\gamma_1+\gamma_2+\ldots+\gamma_{g-1}+P$. Then
for generic $\gamma_1$,\ldots,$\gamma_{g-1}$ one can write an
especially simple formula for the kernel $\omega(\vec t,z,w)$:
\begin{eqnarray}
\label{eq:omega_expl}
\omega(\vec t,z,w) =
\frac{\theta[\sum\limits_j \vec U_jt_j + \vec A(z)- \vec A(w)- \vec
A(\gamma_1)-\ldots- \vec A(\gamma_{g-1})-\vec K)]}{\theta
[\sum\limits_j \vec U_jt_j - \vec A(\gamma_1)-\ldots- \vec
A(\gamma_{g-1})- \vec K]} \times \nonumber
\\
\times\frac{ C \cdot d\mu(w) } {\theta
[ \vec A(z)- \vec A(w)- \vec A(\gamma_1)-\ldots- \vec A(\gamma_{g-1})-
\vec K]}
\cdot \exp\left[i\sum\limits_j t_j \int_w^z \Omega_j \right]
\end{eqnarray}
Here $\Omega_j$ are meromorphic differentials with an unique pole at
the point $P$,
$$
\Omega_j = d(k^j) + \mbox{regular tems}
$$
and zero $a$-periods, $U_j$ denotes the normalized vector of
$b$-periods for $\Omega_j$:
$$
U_j^k = \frac{1}{2\pi}\oint_{b_k} \Omega_j,
$$
$\vec A(\gamma)$ denotes the Abel transform with the starting point
$P$, $\vec K$ is the vector of Riemann constants, $d\mu$ is the
holomorphic differential with the zeroes $\gamma_1$,\ldots,
$\gamma_{g-1}$. Let $\nu$ be a local coordinate near $P$ such, that
$d\mu = d\nu(1+o(1))$. Then the normalization constant $C$ is defined by:
\begin{equation}
\label{eq:omega_norm}
C= \partial_\nu  \raisebox{-4pt}{$\left|\rule{0pt}{12pt}
  \right.$}_{\nu=0} \theta
[-\vec A(v)- \vec A(\gamma_1)-\ldots- \vec A(\gamma_{g-1})-
\vec K].
\end{equation}
\end{lem}
Using standard arguments, one can easily check, that the expression
(\ref{eq:omega_expl}) is single-valued in $\Gamma$, and for generic
$\vec t$ it has the proper poles in $z$ and $w$. Let $z\rightarrow
w$. Then
\begin{equation}
\label{eq:omega_diag}
\omega(\vec t,z,w)\sim \frac{ C \cdot d\mu(w) } {\theta
[ \vec A(z)- \vec A(w)- \vec A(\gamma_1)-\ldots- \vec A(\gamma_{g-1})-
\vec K]}
\end{equation}
For $z$ different from $\gamma_1$, \ldots, $\gamma_g$, $\gamma^*_1$,
\ldots, $\gamma^*_g$ denominator of (\ref{eq:omega_diag}) has a
first-order zero at $w=z$. If $z=\gamma_j$, $j=1,\ldots,g$, then the
denominator vanishes identically.  If $z=\gamma^*_j$,
$j=1,\ldots,g$, then denominator has a second-order pole. Therefore
the zeroes of the numerator coincides with the zeroes of the
denominator's differential, and the residue of (\ref{eq:omega_diag})
at $w=z$ is regular in $\Gamma$. Therefore this residue is constant.
Normalization (\ref{eq:omega_norm}) means, that the residue is equal
to 1 at $z=\infty$. It completes the proof.

It follows from (\ref{eq:omega_expl}) that the Cauchy-Baker-Akhiezer
Kernel $\omega(\vec t,z,w)$ is meromorphic in all $t_j$. Combining
(\ref{eq:omega_expl}) with (\ref{eq:omega_tran}) we obtain
following:
\begin{lem}
\label{lemma:mer_omega} For any divisor $D$ such, that $\Psi_D(x,z)$
is defined for generic $x$, the kernel $\omega_D(x,z,w)$ is
meromorphic in $x$.
\end{lem}
\begin{rem}
Assume, that operators, associated with the curve $\Gamma$ are
strictly periodic in $x$ with  period $T$. Then following formula is
true:
\begin{equation}
\label{eq:omega_per}
\omega(x+T,z,w)=\omega(x,z,w) \cdot e^{i[p(z)-p(w)]x}.
\end{equation}
\end{rem}

Let us derive the orthogonality relation for BA functions treated as
basis in the space of functions in  $x$-space (\ref{eq:orth2}).
Their inner products already were discussed in the work \cite{GOS}.
We have
$$
\int_{-nT}^{nT}\Psi(x,z)\Psi^*(x,w)d\mu dx= i\omega(x,z,w)\Biggr|_{-nT}^{nT}=
$$
$$
=i\omega(0,z,w)\left[e^{i[p(z)-p(w)]nT}-e^{-i[p(z)-p(w)]nT}\right]=
$$
$$
=[p(w)-p(z)]\omega(0,z,w) \int_{-nT}^{nT} e^{i[p(z)-p(w)]x}dx.
$$
Therefore
$$
\lim\limits_{n\rightarrow\infty}\int_{-nT}^{nT}\Psi(x,z)\Psi^*(x,w)d\mu
dx=2\pi [p(w)-p(z)]\omega(0,z,w)\delta(p(z)-p(w)),
$$
and
$$
\lim\limits_{n\rightarrow\infty}\int_{-nT}^{nT}\Psi(x,z)\Psi^*(x,w)d\mu
dx=0 \ \mbox{for} \ \ z\ne w, \ \ z,w\in \kappa_c.
$$
Let $w\rightarrow z$. Substituting (\ref{eq:omega_diag2}) and taking
into account, that the orientation on the canonical contour
$\kappa_c$ is defined by $dp$, we obtain our final result:

{\bf Following Orthogonality Relations for BA functions as a basis
in the $x$-space, are true:
}
$$
(\int_{-\infty}^{\infty}\Psi(x,z)\Psi^*(x,w)dx)d\mu =
$$
$$
=2\pi (p(w)-p(z))\left[\frac{dp(w)}{p(w)-p(z)} +\mbox{regular
    terms}\right] \delta(p(z)-p(w)) = 2\pi\delta(z,w)
$$

\vspace{2cm}

{\bf Appendix 2.} The Hyperelliptic Case. The Periodic Boundary
Conditions.

{\bf The Cauchy-Baker-Akhiezer Kernel.}

In the hyperelliptic case there exists a simple explicit formula for
the Cauchy-Baker-Akhiezer Kernel:
$$
\omega(x,z,w)=i\frac{\Psi(x,z)\Psi_x^*(x,w)-\Psi_x(x,z)\Psi^*(x,w)}{w-z}
d\mu(w)
$$
It is easy to check, that all analytic properties are
fulfilled. Moreover,
$$
\partial_x \omega(x,z,w) =
[\Psi(x,z)\Psi_{xx}^*(x,w)-\Psi_{xx}(x,z)\Psi^*(x,w)]
 \frac{i}{w-z}d\mu=
$$
$$
=[(-z-U(x))\Psi(x,z)\Psi^*(x,w)-(-w-U(w))\Psi(x,z)\Psi^*(x,w)]
 \frac{i}{w-z}d\mu=
$$
$$
= -i \Psi(x,z)\Psi^*(x,w)d\mu.
$$

{\bf The periodic boundary conditions.}

We assume that our finite-gap operators are periodic with the period
$T$. In addition to the spectral problem in the whole line one can
consider the periodic boundary problem with an fixed unitary
multiplier:
\begin{equation}
\Psi(x+T,z)=\varkappa\Psi(x,z),  \ \ |\varkappa|=1.
\end{equation}
For regular potentials this problem is self-adjoint and has only
discrete spectrum. Let us enumerate the points $z_j$ in $\Gamma$,
$j=1,2,\ldots,\infty$ such, that
$$
e^{iTp(z_j)}=\varkappa.
$$
All these points lie in the canonical contour $\kappa_0$. Each finite
oval contains only finite number of points $z_j$.
\begin{lem}
The scalar product for the basic eigenfunctions is given by:
\begin{equation}
\int_0^T \Psi(x,z_j)\Psi^*(x,z_k)dx = \delta_{jk}\frac{dp(z_j)}{d\mu(z_j)}.
\end{equation}
As above, we deform the integration contour in the $x$-plane to
avoid singularities.
\end{lem}
Let us point out that in the  real self-adjoint case with positive
inner product this formula was known--see the work \cite{Kr},
formula 30. Our lemma extends this result to the case of singular
potentials and indefinite inner product.

 {\bf Proof.} For $j\ne k$
$$
\int_0^T \Psi(x,z_j)\Psi^*(x,z_k)dx = \frac{i\omega(x,z_j,z_k)}{d\mu(z_k)}
\Biggr|_{0}^{T}=
\frac{i\omega(0,z_j,z_k)}{d\mu(z_k)}\left[e^{iT[p(z_j)-p(z_k)]}-1  \right] =0.
$$
Let $k=j$.
$$
\int_0^T \Psi(x,z_j)\Psi^*(x,z_j)dx = \lim\limits_{w\rightarrow z_j}
\frac{i\omega(x,z_j,w)}{d\mu(w)}\Biggr|_{0}^{T}=
$$
$$
=\lim\limits_{w\rightarrow z_j} \left[
  \frac{idp(w)}{d\mu(w)[p(w)-p(z_j)]} +\mbox{regular terms} \right]
\left[e^{iT[p(z_j)-p(w)]}-1  \right] = \frac{dp(z_j)}{d\mu(z_j)}.
$$
Assume now, that the spectral data $\Gamma$, $D$ satisfy the reality
constraints.  Taking into account that for real case
$\Psi^*(x,z)=\bar\Psi(x,\tau z)$, we obtain following:

{\bf Theorem.}
{\it Let us define the scalar product by
\begin{equation}
\label{eq:orth3}
(\Psi(x,z_j),\Psi(x,z_k))_{x}=\int_0^T\Psi(x,z_j)\bar\Psi(\bar x,z_k).
dx
\end{equation}
For singular potentials this scalar product is indefinite. The
dimension of negative subspace is finite and coincides with the
number of points $z_j$ such, that $dp(z_j)/d\mu(z_j)<0$. Therefore
we have a {\bf Pontryagin-Sobolev} space of functions where our
singular finite-gap operator is symmetric}.

{\bf Example: The Higher Lame Potentials}.  Let us consider the
operators $L=-\partial^2+n(n+1)\wp(x)$ for the semi-canonical case
where elliptic function is defined by the Riemann Surface
$y^2=(u-u_0)(u-u_1)(u-u_2), u_j\in R$. For $n=1$ this case was
discussed in Chapter 2. For all $n\in Z^+$ these potential are
singular finite gap. They  correspond to some algebraic curve
$\Gamma_n$ with genus $g=n$ and poles in the points $mT,m\in Z$.
Here T is a real period of the function $\wp(x)$. The operator
$L_{n,i\omega}=-\partial^2+n(n+1)\wp(x+i\omega)$ is smooth
finite-gap for the imaginary period $2i\omega$. It corresponds to
the same algebraic curve $\Gamma_n$.

 {\bf
The Hermit Spectrum here also coincides with the set of real points
$S_n(u)=\pm 1, S_n'(u)=0$ as for $n=1$. Here $S_n=1/2Tr\hat{T}_n$ is
the trace on monodromy matrix for the operators $L_n$ or
$L_{n,i\omega}$ on the real line.} For the case $L_n$ our divisor is
equal to $D=n\infty$. For the operator $L_{n,i\omega}$ with even
nonsingular periodic potential, our divisor is located in the upper
boundaries of the finite gaps of spectrum. They   are the branching
points $u^*_{n,2k},k=1,2,...,2n$ for the algebraic curve $\Gamma_n$.
All Hermit Spectrum here is located in the infinite spectral zone
$[u^*_{n,2n},\infty]$ as for $n=1$ (see chapter 2).

Let us construct the space $H_L$ for the Lame operators $L=L_n$.
Elementary local analysis with Laurent series near the pole $x=0$
leads to the following result: {\bf All eigenfunctions of the
operator $L_n$ have following singularities}:
$$n=2k-1:\psi(x)=\alpha_1/x^{n}+\alpha_2/x^{n-2}+...+\alpha_k/x+\phi(x)$$
$$n=2k:\psi(x)=\alpha_1/x^n+\alpha_2/x^{n-2}+...+\alpha_k/x^2+\phi(x)$$
where $$\phi(0)=...=\phi^{(2r)}(0)=...=\phi^{(2k-2)}(0)=0$$ for
$n=2k-1$, and $$\phi'(0)=...\phi^{2r}(0)=...=\phi^{2k-2}(0)$$ for
$n=2k$\footnote{Local analytic properties of wave functions for
singular potential with such singularities in case $n=1$ were
discussed in  \cite{APP}.} Therefore all their linear combinations
also satisfy to these linear relations. We are coming to the
following result:

{\it Consider the spaces of $C^{\infty}$-functions $f(x)\in
F_{n,\varkappa}, f(x+T)=\varkappa f(x)$, and
$$f(x)-f(-x)=O(x^n), n=2k$$
 $$f(x)+f(-x)=O(x^n), n=2k-1$$
 We introduce in these spaces the standard positive inner product.
Consider the spaces of function $F_{n,\varkappa}\bigoplus
C^k(\varkappa)$ where the space $C^k(\varkappa)$ over $C$ is
generated by the singular BA eigenfunctions with negative inner
squares defined above, with Bloch-Floquet multiplier $\varkappa$. We
have $k=[(n+1)/2]$ so all this extension exactly corresponds to the
set of ''residues'' $\alpha_1,...,\alpha_k$. They have singularities
with Laurent part described here. The direct integral of these
spaces over the circle $\varkappa\in S^1$ is our space
$\tilde{H}_{L_n}$ with indefinite metric.}

Construct KdV solution with initial value $U(x,0)=n(n+1)\wp(x)$
where $$U_t=6UU_x-U_{xxx}$$ It has a form like
$$U(x,t)=\sum_{j=1}^{j=n(n+1)/2}2\wp(x-x_j(t))$$
such that $x_j(0)=0$. We assume that there exists and unique
solution such that $x_j\neq x_p$ for small $t$.

{\bf Statement}. We have asymptotically $x_j\sim a_jt^{1/3}$ for
$t\rightarrow +0$ and $a_j\neq a_p$. Here $a_j$ in general are
complex numbers. This set is invariant under transformation
$a_j\rightarrow \eta a_j$ where $\eta^3=1$. Every orbit of the group
$Z_3$ generated by $\eta$, enter this set only ones. This group acts
freely if $n(n+1)/2$ is divisible by 3. There is exactly one zero
orbit $a_{n(n+1)/2}=0$ if $n(n+1)/2=1 (mod 3)$. Other orbits are
free and distinct, i.e. $a_j^3\neq a_k^3$ if they belong to
different orbits.

In order to prove that, we consider rational solutions
$U_0=\sum_j2/(x-x_j(t))^2$ such that $x_j(0)=0$. For the poles we
have an equation $$\partial_tx_j(t)=-12\sum_{p\neq j}1/(x_j-x_p)^2$$
Changing time, we remove  inessential factor $-12$ keeping the same
notation for time. The solution $U_0$ approximates $U$ near the pole
$t\rightarrow +0, x\rightarrow 0$. Our functions $x_j=a_jt^{1/3}$
exactly satisfy it. So we obtain algebraic equation for the set
$a^3_j$. Our statement easily follows from this algebraic equation.

{\bf Conjecture.} There exists exactly $k=[(n+1)/2]$ ''real'' orbits
(i.e. $a_j^3\in R$).

Let us point out that rational solutions to the KdV hierarchy with
initial function $U_0(x,0)=n(n+1)/x^2$ were investigated in the work
\cite{AM}. Some general formulas were obtained for this solutions.
In particular, they proved that at least one pole is real (because
corresponding Lax Operator is not self-adjoint).

According to our work,  this number is exactly equal to the number
of negative squares in the indefinite inner product for every value
of $\varkappa$. It is obvious that the number of real orbits is
$k'\geq k=[(n+1)/2]$ but opposite inequality should be proved. Our
space $\tilde{H}_{L_t}$ for the operators $L_t=-\partial_x^2+U(x,t)$
can be obtained by the extension of the space of
$C^{\infty}$-functions with zeroes in the poles $x_j(t)$. So the
dimension of additional part is exactly equal to the number of poles
if and only if there are no relations between residues. The
dimension of negative part for $L_t$ is stable under KdV deformation
(i.e.remains constant).

{\bf We proved this conjecture numerically for $n<9$.}

 For $n=1$ we have
$a_1=0$.

 For $n=2$ we have $a_j=(1,\eta,\eta^2)$.

  For $n=3$ we have
$a_j=(a,a\eta,a\eta^2,b, b\eta,b\eta^2)$. Here we have
$$w=a^3/b^3=1/2(-7+\sqrt{45})$$ and both orbits are real
$$a^3=1+9w(w+2)/(w-1)^2,b^3=1+9w^{-1}(w^{-1}+2)/(w^{-1}-1)^2$$

As a Corollary, we obtain an important result: {\bf There are no
relations between residues of the real poles  for the generic
finite-gap real singular potential in this specific case. So  the
space $\tilde{H}_{L_t}$  of functions in the variable $x$ for
$L=-\partial_x^2+U(x,t)$ depends on the poles only.}

For Higher Lame Potentials we have {\bf ''The Intermediate Cases''}
$$D=r\infty +\gamma_1+...+\gamma_{n-r}$$ where $n=g$ and all points
$\gamma_j=(\lambda^*_j,\pm)$ are located in some finite gaps (no
more than one point in one finite gap).  The Hermit Spectrum
contains the same points as for higher Lame potential: they were
described through the half-trace $S(u)$ of monodromy matrix
$S=1/2Tr\hat{T_n}$ as $S=\pm 1, S'=0$. However, here the Hermit
Spectrum contains also finite number of points $\lambda^*_j$ coming
from the divisor points in finite gaps. These additional spectral
points $\lambda^*_j$ can be obtained using some carefully chosen
chain of ,,Backlund-Darboux Transformations'' starting from the
smooth potential $L_{n,i\omega}$: sometimes  they simply lead to the
special shifts along the Jacoby Variety in the theory of finite-gap
 potentials.

{\bf Generalizations: The Intermediate and One Pole finite-gap
singular potentials}.

{\bf The Intermediate Potentials}. Consider any  nonsingular
hyperelliptic curve $\Gamma$ given as $y^2=(u-u_0)...(u-u_{2n})$
with real $u_0<...<u_{2n}$. We chose divisor
$D=r\infty+\gamma_1+...+\gamma_{n-r}$ where all finite points
$\gamma_j=(\lambda^*_j, \pm )$ belong to some finite gaps
$a_k=[u_{2k-1}u_{2k}], k\geq 1$, (no more than one point in the
gap). Let us assume that this potential is $x$-periodic for real
$x$, with minimal period $T$. Corresponding potential $U_(x)$ has
some number of poles on the circle $0T$, of the form
$r(r+1)/x^2+O(x)$ each. We can prove that the Hermit spectrum in
this case consists of all points $S(u)=\pm 1, S'=0$, plus all points
$\lambda^*_j$ in the finite gaps. Minimal set of Hermit eigenvalues
we have for the case $r=n$. Maximal set we have for $r=0$, i.e. for
the nonsingular finite-gap potentials.

 Other potentials of that
family can be obtained from the nonsingular one with $r=0$  by the
special sequence of Backlund-Darboux Transformations ($\Gamma$ is
fixed).

No problem to describe here (for all $r\leq n$) the number of
eigenvalues with negative inner square for  periodic spectrum: Put
first the sign $+$ on the infinite spectral zone and move to the
next one from the left. Put sign $-$ on the next spectral zone if
there is no divisor points between them. Otherwise put the same sign
on the next zone. After that continue to move left with the same
rule. So we have exactly $r$ changes of sign. It is equal to the
number of finite gaps empty from the divisor points. Additional
negative periodic singular eigenfunction is coming from the points
$S(u)=1, S'=0$ inside of finite spectral zones with the negative
sign. We can think about that point as ''degenerate gap'' of the
length zero with additional divisor point in it, so it does not
change sign. So the maximal number of negative signs we obtain for
the case $r=1$ where the empty gap is exactly the maximal finite gap
$[u_{2n-1}u_{2n}]$. However, the detailed description of indefinite
functional space in the $x$ variable depends here on the number of
poles which is hard to predict.

{\bf  Construct the Infinite Gap limit with fixed $r$ equal to the
number of empty finite gaps.}  This class of potentials might be
especially interesting. We can introduce the infinite-gap limit of
them: We keep the number $r$ with corresponding empty finite gaps,
real period $T$ and the form and number of poles unchanged for
$n\rightarrow\infty$, opening more and more small gaps (branching
points) with new divisor points
 inside of infinite spectral zone. We open them near the points
$S(u)=\pm 1$ such that their length tends to zero rapidly enough.
Probably, it is possible to construct such singular infinite-gap
periodic potentials simply by the sequences of specially chosen
Backlund-Darboux transformations from the smooth periodic
potentials. We need to prove that all such ''infinite-gap''
intermediate potentials can be approximated by the singular
finite-gap potentials with the same period and fixed $r$.

 {\bf The
One Pole  Potentials}. Consider the special case $D=n\infty$
periodic in $x$ with minimal period $T$. Let corresponding operator
(i.e. its algebraic curve $\Gamma$ and period $T$) is  small enough
perturbation of the Lame potential $n(n+1)\wp(x)$. In this case the
potential $U(x)$  has only one real pole on the circle $0T$ with
singularity $n(n+1)/x^2+O(x^2)$. All results formulated above for
Lame operators are valid here (some of the them require evenness of
the perturbed potential).

 More general one-pole potentials can be constructed:
  Let $U(x)$ be any smooth periodic
finite-gap potential with minimal period $T$. Following \cite{Crum},
one can apply a special Backlund (i.e. Darboux) transformations (the
Crum transformations) to $U(x)$. The resulting potential $U_1(x)$ is
also periodic with the same period. It has the same spectral curve
$\Gamma$ and has exactly one real pole on the circle $0T$ at the
point $0$ with singularity $2/x^2+O(x)$. The Dirichlet spectrum
(Hermit spectrum in our paper) of $U_1(x)$ is obtained from the
Dirichlet-Hermit spectrum of $U(x)=U_0$ by removing the ground
state. One can iterate this procedure $U_j\rightarrow U_{j+1}$,
removing the ground state at each step and increasing the leading
coefficient of the pole at the point $0$ from $j(j+1)/x^2+O(x)$ to
$(j+1)(j+2)/x^2+O(x)$.

{\bf Statement.} Denote by $U_r(x)$ the potential, obtained from the
smooth finite-gap potential $U(x)=U_0$ by $r$ iterations of the Crum
transformation over the ground state functions. Then $U_r(x)$ is
periodic with  period $T$ and has exactly one pole of type
$r(r+1)/x^2+O(x)$ at the point $x=0$. The spectral curve of $U_r(x)$
coincides with the spectral curve for $U(x)=U_0$, and the
Hermit-Dirichlet spectrum of $U_r(x)$ coincides with the
Hermit-Dirichlet spectrum of $U(x)$ with $r$ lowest states removed.

This procedure can also be applied to smooth periodic infinite-gap
potentials with minimal period $T$.

{\bf Examples: Degenerate Cases.} Consider now the degenerate
singular one-gap potentials corresponding to the degenerations of
spectral elliptic curves associated with Weierstrass elliptic
functions and Lame operators:

a. Trigonometric   (hyperbolic) singular degeneration:

The real period tends to infinity, and imaginary period remains
finite
$$2\wp(x)+const\rightarrow 2c^2/\sinh^2(cx)=U_c(x)$$ Here elliptic
functions correspond to the algebraic curve $\Gamma_t, t\rightarrow
0$. Realizing it as a spectral curve of our operators, we obtain
branching points $u_0^t<u_1^t<u_2^t<\infty$ such that $u_j^t\in R$.
In process of degeneration $t\rightarrow 0$ we have
$$u_0,u_1\rightarrow -c^2, u_2\rightarrow 0$$ The divisor
point $\gamma$ coincides with $\infty$ for all $t$. Easy to see that
the spectrum of  operator $L_c=-\partial_x^2+U_c(x)$ in the
corresponding space $\tilde{H}_{L_c}$ on the whole line $R$ occupies
all set $\lambda\geq 0$ plus one negative discrete eigenvalue. We
have singular eigenfunction
$$\psi_c=1/\sinh(cx)$$ for $\lambda_0=-c^2$. Our spectrum  consists of the
union of the half-line
 $\lambda\geq 0$ and one-dimensional space for $\lambda_0=-c^2$. We
    have exactly one negative simple eigenvalue
$-c^2=\lambda_0<0$. It has negative inner square in the
Pontryagin-Sobolev space constructed above. All other eigenvalues
are positive. The number of negative squares on the whole line is
equal to one.

b. Rational degeneration $c\rightarrow 0$: Here $U\rightarrow
2/x^2$. Both periods became infinite, and $u_j^t\rightarrow 0,
j=0,1,2$. The spectrum of operator $L=-\partial_x^2+2/x^2$ on the
whole line became $\lambda\geq 0$. For $\lambda=0$ we have singular
eigenfunction $\psi_0=1/x$ (the discrete eigenvalue on the bottom of
the continuous spectrum) whose inner square is negative in our inner
product.

c. Easy to consider also trigonometric degeneration
$U=2c^2/\sin^2(cx)$ with  antiperiodic eigenstate corresponding to
the endpoint of limiting ''finite spectral zone'' $0\leq \lambda\geq
c^2$,
$$\psi_0=1/\sin(cx), L_c\psi_0=c^2\psi_0$$
Infinite spectral zone also starts in the same point $\lambda\geq
c^2$. Here real period remains finite and imaginary period tends to
zero. This case corresponds to the degeneration of real elliptic
curves corresponding to the zero length of finite gap $u_1=u_2=c^2$
and $u_0=0$. The divisor point is equal to $\infty$. We have
spectrum $\lambda\geq u_0$. For every Bloch-Floquet multiplier
$\varkappa\in S^1$ we have exactly one singular eigenfunction with
eigenvalue $0\leq \lambda_0(\varkappa)\leq c^2$ whose inner square
is negative in our indefinite metric. The groundstate
$\phi_0=\cos(cx)/\sin(cx)$ for $\lambda=0$ corresponds to the
periodic case $\varkappa=1$. This spectrum can be obtained from the
constant potential $U=0$ by the Crum-Darboux Transformation
discussed above (here $c=1$):
$$L\rightarrow \tilde{L}, L=-(\partial_x+a)(\partial_x-a)=-\partial_x^2+U,L\eta=\lambda'\eta$$
$$\tilde{L}=-(\partial_x-a)(\partial_x+a)=-\partial_x^2+\tilde{U}, a=(\log \eta)_x$$
$$[Lf=\mu f,\tilde{f}=(\partial_x-a)f]\rightarrow
[\tilde{L}\tilde{f}=\mu\tilde{f}]$$
$$\tilde{U}=U-2a_{x}=U-2(\log \eta)_{xx}$$
Taking here $U=0,\lambda'=1,\eta=\sin (x)$, we obtain
$$\tilde{U}=2/\sin^2(x),[\mu=0,f= 1]\rightarrow \tilde{f}=\phi_0=\cos(x)/\sin(x)$$
and
$$[\mu=1,f=\cos(x)]\rightarrow [\tilde{f}=(\partial_x-a)\cos(x)=-1/\sin(x)]$$

\end{document}